\documentclass{iau} 
\usepackage{graphicx}
\usepackage{hyperref}
\usepackage{natbib}
\usepackage{multirow}
\usepackage[T1]{fontenc}
\usepackage{array}
\usepackage[table]{xcolor}
\usepackage{amsmath}
\usepackage{graphicx}
\usepackage{subfig}
\usepackage[colorinlistoftodos]{todonotes}
\usepackage{natbib}

\newcolumntype{s}{>{\columncolor[HTML]{DAF7A6}} c}
\pubyear{2022}
\volume{368}
\jname{Machine Learning in Astronomy: Possibilities and Pitfalls}
\editors{J.~McIver, A.~Mahabal \&, C.~Fluke eds.}

\title[Stellar classification through machine learning] 
{Spectral identification and classification of dusty stellar sources using spectroscopic and multiwavelength observations through machine learning}
\author[Sepideh Ghaziasgar et al.]   
{Sepideh Ghaziasgar$^1$, Amirhossein Masoudnezhad$^2$, Atefeh Javadi$^1$, Jacco Th. van Loon$^3$, Habib G. Khosroshahi$^1$, and Negin Khosravaninezhad$^2$\ \let\comma,$^4$}

\affiliation{$^1$School of Astronomy, Institute for Research in Fundamental Sciences (IPM), P.O. Box 19395-5531, Tehran, Iran  \\ [\affilskip]
	
$^2$Department of Physics, Sharif University of Technology, P.O. Box \\ 11155-9161, Tehran, Iran \\ [\affilskip] $^3$Astrophysics Group, Lennard-Jones Laboratories, Keele University, ST5 5BG, UK \\ [\affilskip] $^4$Department of Physics and Astronomy, University of California Riverside, CA 92521, USA}

\begin{document}

\maketitle

\begin{abstract}
We proposed a machine learning approach to identify and distinguish dusty stellar sources employing supervised and unsupervised methods and categorizing point sources, mainly evolved stars, using photometric and spectroscopic data collected over the IR sky. Spectroscopic data is typically used to identify specific infrared sources. However, our goal is to determine how well these sources can be identified using multiwavelength data. Consequently, we developed a robust training set of spectra of conﬁrmed sources from the Large and Small Magellanic Clouds derived from SAGE-Spec Spitzer Legacy and SMC-Spec Spitzer Infrared Spectrograph (IRS) spectral catalogs. Subsequently, we applied various learning classifiers to distinguish stellar subcategories comprising young stellar objects (YSOs), C-rich asymptotic giant branch (CAGB), O-rich AGB stars (OAGB), Red supergiant (RSG), and post-AGB stars. We have classified around 700 counts of these sources. It should be highlighted that despite utilizing the limited spectroscopic data we trained, the accuracy and models' learning curve provided outstanding results for some of the models. Therefore, the Support Vector Classifier (SVC) is the most accurate classifier for this limited dataset.

\end{abstract}
\firstsection 
\section{Introduction}              
Surveys have acquired large datasets over the multiwavelength sky. Understanding dusty stellar sources, such as young stellar objects (YSOs) and evolved stars, has been dramatically enhanced by the acquisition and analysis of spectroscopic data. Indeed, the growing number of spectroscopic observations and confirmed stellar spectra has necessitated the development of novel approaches for identifying and classifying stellar sources. As a result, machine learning enhances the accuracy of spectral classification of dusty point sources while also validating candidate spectroscopic catalogs and identifying unclassified spectra.\

For probing the Magellanic Clouds, photometric and spectroscopic classification of star properties and content is used. Although photometric approaches are faster, spectroscopic methods provide a more accurate analysis of large samples. Indeed, the Magellanic Clouds are suited for investigating the stellar contribution to dust emission (Ruffle et al. 2015). An early-evolutionary star is referred to as a YSOs. Due to star formation, YSOs are surrounded by gas and dust disks (Suh 2016). Therefore, YSOs contribute to our understanding of star formation and allow us to characterize the star-forming regions. Mass, age, and starforming environment determine the properties of YSOs. In addition, evolved stars are dust producers, and their role in enriching galaxies is crucial.\

In the final stage of stellar evolution, low– and intermediate mass (0.8-8 \(\textup{M}_\odot\)) stars evolve into the Asymptotic Giant Branch (AGB) phase, and high mass (M$\ge$ 8 \(\textup{M}_\odot\)) stars enter the Red Supergiants (RSG) phase. Most AGB stars are long-period variables (LPVs) with large amplitude pulsations; and they have circumstellar dust envelopes with high mass-loss rates (Saremi et al. 2020; Navabi et al. 2021; Javadi \& van Loon 2022). The dust production and rates of mass loss of pulsating AGB stars and RSG stars are determined by evolved stars. AGB stars are classified as O-rich AGB (OAGB) or C-rich AGB (C-AGB) based on the chemistry of the photosphere and the outer dust envelope (Suh 2020, 2021; Javadi et al. 2011a, 2011b, 2013). We can detect these sources in the infrared band due to AGB stars and RSGs are luminous $(\sim 10^{3.5-5.5}\  \textup{L}_\odot\)) and red (Javadi \& van Loon 2022). The post-AGB stars are luminous, low- or intermediate-mass stars, which end their AGB branch evolution by a phase of mass loss (Van Winckel 2003). Consequently, infrared (IR) spectral observations and classification are essential for studying YSOs and evolved stars (Boyer et al. 2017).

\section{Data}
We used tabular FITS datasets available from the Strasbourg astronomical Data Center (CDS) for the Large and Small Magellanic Clouds taken from SAGE-Spec Spitzer Legacy and SMC-Spec Spitzer Infrared Spectrograph (IRS) spectral catalogs (Jones et al. 2017a; Ruffle et al. 2015). SAGE is the project for Surveying the Agents of Galaxy Evolution for tracking dust and gas in  Magellanic Clouds (Meixner et al. 2006; Woods et al. 2011; Ruffle et al. 2015; Jones et al. 2017a). To prepare the final data for implementation on various models, we performed some preprocessing, including label encoding and data imputation. Following that, we selected 22 features including magnitude and flux at different wavelengths, and 694 samples were collected for both catalogs. Consequently, we allocated a feature data frame (X) with 694 rows and 22 columns and a target array (Y) for each sample's related class. We used some features comprising Umag, Bmag, Vmag, Imag, <Vmag>, <Imag>, J2mag, H2mag, Ks2mag, Jmag, Hmag, Ksmag, 3.6$\ \mu m$, 4.5$\ \mu m$, 5.8$\ \mu m$, 8.0$\ \mu m$. Stellar classes with total counts are listed in Table 1.\\\\

 \begin{table}[h!]
 \centering
 \begin{tabular}{c  c c  c}
 \hline\\
 Classes & LMC & SMC & Total \\ [0.8ex] 
 \hline \\
 YSO & 213 & 51 & 264 \\ 
 C-AGB & 152 & 39 & 191 \\
 O-AGB & 89 & 19 & 108 \\
 RSG & 72 & 22 & 94 \\
 Post-AGB & 33 & 4 & 37 \\ [0.8ex] 
 \hline	
 \end{tabular}
 \caption{Number of samples}
 \label{table:1}
 \end{table}


\section{Method}

Machine Learning (ML) deals with techniques and methods that enable the machine to learn. In this regard, algorithms are developed to perform tasks and construct models to make inferences based on the data type. Moreover, these inferences can be a pattern, description, or prediction based on past data (Sturrock et al. 2019). Training, validation, and testing datasets are commonly used in model development. Machine learning methods are applied to astronomy thanks to big astronomical data from the ground-based and space telescopes. Classification is one of the areas where astronomers are interested in implementing supervised and unsupervised algorithms ({Ivezi\'{c} et al. 2014). To categorize dusty stellar sources, we used K-Nearest Neighbor (KNN), Decision Tree Classifiers (DTC), Linear Discriminant Analysis (LDA), Gaussian Naive Bayes, Stochastic Gradient Descent (SGD), and Support Vector Classification (SVC) for limited spectral data.\ 

To understand some results in the following, some of the standard definitions and functions should be explained briefly.\

\textit{Classification report}; contains the model's precision, recall, and f1-score values for each class. Precision; is the fraction of relevant instances among the retrieved ones. Recall; is the fraction of relevant cases retrieved, and in the following formula, f1-score is a combination of precision and recall.\[
f_1\text{-}score = \frac{2\times precision\times recall}{precision+recall} \] 

\textit{Accuracy}; As a metric for evaluating the classification models is, the number of correct predictions divided by the total number of predictions.\

\textit{Learning Curve Function}; This function splits the dataset into training and test data. It determines the cross-validated training and test scores to demonstrate how the performance of the classifier changes with varying training samples. The learning curve optimizes algorithms and their parameter over time as they learn gradually.\

\textit{Validation Curve}; This curve determines and plots training and test scores for different values of a hyperparameter of the classifier.

\textit{Confusion Matrix}; This is an N x N table layout that visualizes an algorithm's performance in which rows represent instances in the actual class, and the columns represent the predicted class. 

\section{Some Results and On-going work}

As shown in Table 2, we implemented several models. For example, in the following paragraph, we presented the Support Vector Classifier (SVC) as a supervised learning method for classification, which works by mapping data to a high-dimensional feature space so that data points can be classified even when the data are not otherwise linearly separable (Kovács, Szapudi 2015; Krakowski et al. 2016).\

 In this study, we tuned the values of SVC hyperparameters (e.g., C, gamma), and based on validation curves for hyperparameters, C should be more than 100 and gamma $\sim$ 0.1, respectively (Figure 3). The final results showed 90 percent accuracy and 88 percent f1-score average, which were the best among all models. Therefore, SVC is the most accurate classifier for our dataset without augmentation, as shown in table 2. The confusion matrix has shown the SVC model's performance for stellar classifications (Figure 1).\
 
In the future, since the lack of data and its imbalance resulted in lower classification scores in some classes (e.g., PostAGBs), we will apply augmentation to our dataset to improve our accuracy. We are looking for an optimal model for our data augmentation and applying it to classifiers to demonstrate the models’ performance before and after augmentation. One way to solve this problem is to oversample the minority class data, which can be achieved by synthesizing samples from the minority classes before fitting a model.\
To further investigate, we aim to complete this study for machine learning models to categorize dusty stars that can improve by employing data augmentation and expanding spectroscopic data and approved catalogs, including the JWST and upcoming spectroscopic survey data. A mid-infrared stellar population study of galaxies is generally limited to nearby galaxies, whereas JWST will allow studies of galaxies within the Local Volume (Jones et al. 2017b). Additionally, we will be able to enhance the functionality and accuracy of our ML models by training sophisticated classifiers.\\
\

\begin{table*}[ht]
\centering
\resizebox{0.8\textwidth}{!}{
	\begin{tabular}{c c  c c}
	\hline 
	
	\multirow{2}{*}{Classifier} &  & 
	\\ 
	&  &  \ Average f1-Score (\%) & \ Accuracy (\%)  \\ [0.8ex] 
	\hline 
	\\
	K-Nearest Neighbors & & 79 & 86 \\ \hline
	\\
	Decision Tree Classifier & & 80 & 86 \\ \hline
	\\
	Linear Discriminant Analysis & & 86 & 89  \\ \hline
	\\
	Gaussian Naive Bayes& & 77 & 81  \\ \hline	
	\\
	
	Stochastic Gradient Descent & & 87 & 90 \\ \hline
	\\
	Support Vector Classification & & 88 & 90 \\		
	\hline	
	\	
	\end{tabular}}

\caption{Performance of Classifiers: This table compares the performance of different classification models based on f1-Score (\%) and accuracy (\%). The Support Vector Classifier (SVC) is the most accurate classifier for this limited dataset with 90 percent accuracy and an average f1-score of 88 percent.}

\label{table:3}\
\end{table*}


\begin{table}[h!]
\centering
\resizebox{0.58\textwidth}{!}{%
	\begin{tabular}{c c c c c}
	\hline \\
	& precision  &  recall & f1-score  & support \\ [0.9ex] 
	\hline \\
	CAGB & 0.95 & 1.00 & 0.97 & 37 \\
	OAGB & 0.80 & 0.77 & 0.78 & 26 \\
	PAGB & 0.88 & 0.88 & 0.88 & 8 \\
	RSG   & 0.88 & 0.75 & 0.81 & 20 \\
	YSO   & 0.92 & 0.96 & 0.94 & 48 \\
	\hline 
	\\	accuracy &    &     &         0.90 & 139 \\
	macro avg & 0.89 & 0.87 & 0.88 & 139 \\
	weighted avg & 0.90 & 0.90 & 0.90 & 139 \\ [0.9ex]
	\hline
	\
\end{tabular}}

\caption{In the classification report, the f1 score is calculated by averaging. The macro avg (average) f1 score represents an average of the f1 scores over classes. The weighted avg (average) f1 score is calculated by the mean of all per-class f1 scores while considering each class’s support. Support is determined by the number of occurrences of the category.}

\label{table:3}
\end{table}

\begin{figure}[h!]
\centering 
\includegraphics[width=0.69\textwidth]{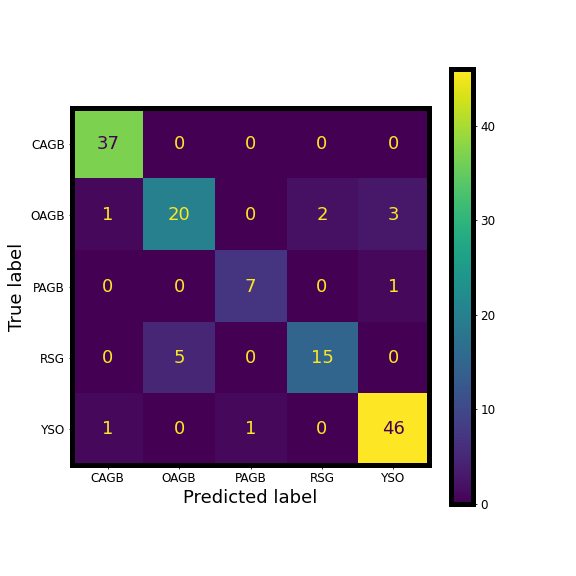} 
\caption{Confusion Matrix of the Support Vector Classifier: This layout shows how well a classification algorithm performs. Based on the model, it predicts that a certain number of objects belong to each class. The dusty stellar class diagonals show the performance of predicted and true labels.} 	
\end{figure} 

\begin{figure}[h!]
\centering
\includegraphics[width=0.9\textwidth]{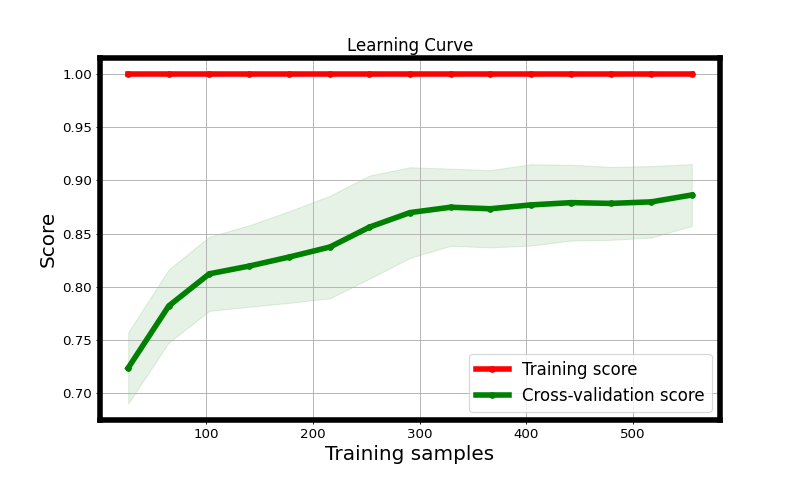}
\caption{Learning Curve of the Support Vector Classifier: This curve shows that learning increases gradually and proper training samples fit the model.}  
\end{figure}

\pagebreak

\begin{figure}
\centering
\subfloat[]{
	\includegraphics[width=0.52\textwidth]{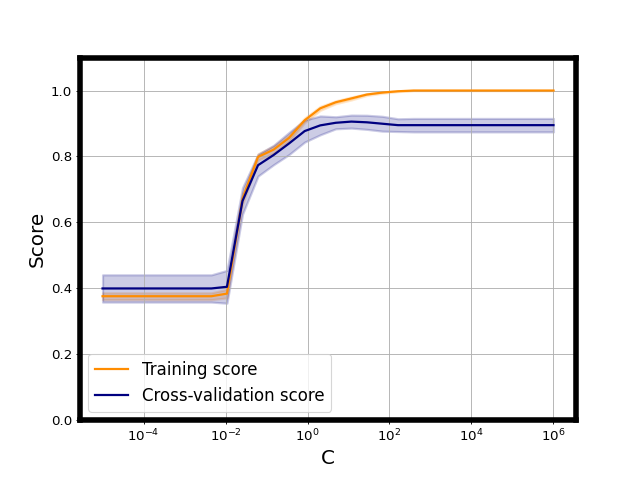}\label{fig1a}} 
\subfloat[]{
	
	\includegraphics[width=0.52\textwidth]{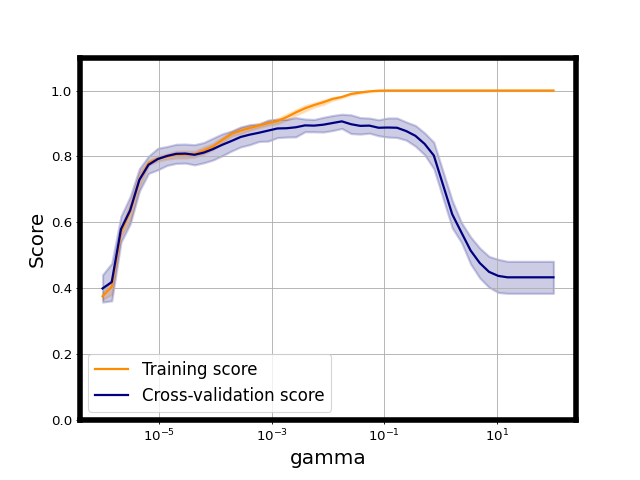}\label{fig1b}}
\caption{Validation Curve of the Support Vector Classifier for the hyperparameters of C and gamma. These SVC hyperparameter values are tuned based on validation curves. As shown, C is more than 100 and gamma $\sim$ 0.1.}\label{fig2} 
\end{figure}




\pagebreak
\def\apj{{ApJ}}    
\def\nat{{Nature}}    
\def\jgr{{JGR}}    
\def\apjl{{ApJ Letters}}    
\def\aap{{A\&A}}   
\def\mnras{{MNRAS}}
\def\aj{{AJ}}
\let\mnrasl=\mnras

\pagebreak 

\
\section*{Acknowledgements}
The authors thank the School of Astronomy at the Institute for Research in Fundamental Sciences (IPM) and the Iranian National Observatory (INO) for supporting this research. S. Ghaziasgar is grateful for the support from the IAU.



\end{document}